\documentstyle[12pt,axodraw,graphicx]{article}
\begin{document}
\baselineskip 0.6cm

\def\simgt{\mathrel{\lower2.5pt\vbox{\lineskip=0pt\baselineskip=0pt
           \hbox{$>$}\hbox{$\sim$}}}}
\def\simlt{\mathrel{\lower2.5pt\vbox{\lineskip=0pt\baselineskip=0pt
           \hbox{$<$}\hbox{$\sim$}}}}

\begin{titlepage}

\begin{flushright}
UCB-PTH 04/06 \\
LBNL-54674 \\
\end{flushright}

\vskip 2.0cm

\begin{center}

{\Large \bf 
Warped Supersymmetric Unification with Non-Unified Superparticle Spectrum
}

\vskip 1.0cm

{\large
Yasunori Nomura$^{a,b}$, David Tucker-Smith$^c$, and Brock Tweedie$^{a,b}$
}

\vskip 0.4cm

$^a$ {\it Department of Physics, University of California,
                Berkeley, CA 94720} \\
$^b$ {\it Theoretical Physics Group, Lawrence Berkeley National Laboratory,
                Berkeley, CA 94720} \\
$^c$ {\it Department of Physics, Williams College, 
                Williamstown, MA 01267}

\vskip 1.2cm

\abstract{We present a new supersymmetric extension of the standard model. 
 The model is constructed in warped space, with a unified bulk symmetry 
 broken by boundary conditions on both the Planck and TeV branes.  In 
 the supersymmetric limit, the massless spectrum contains exotic colored 
 particles along with the particle content of the minimal supersymmetric 
 standard model (MSSM). Nevertheless, the model still reproduces the MSSM 
 prediction for gauge coupling unification and does not suffer from a 
 proton decay problem. The exotic states acquire masses from supersymmetry 
 breaking, making the model completely viable, but there is still the 
 possibility that these states will be detected at the LHC.  The lightest 
 of these states is most likely $A_5^{\rm XY}$, the fifth component of 
 the gauge field associated with the broken unified symmetry.  Because 
 supersymmetry is broken on the $SU(5)$-violating TeV brane, the gaugino 
 masses generated at the TeV scale are completely independent of one 
 another.  We explore some of the unusual features that the superparticle 
 spectrum might have as a consequence.}

\end{center}
\end{titlepage}

\section{Introduction}

Because it stabilizes the weak scale and leads naturally to gauge coupling 
unification, weak-scale supersymmetry has attracted much interest.  For 
many, the unification of couplings suggests that some new unified physics 
emerges at the scale at which the couplings meet~\cite{Dimopoulos:1981zb}.  
But this scale is enormously large, and so our experimental probes of it 
are quite limited: we must rely on searches for extremely rare processes 
such as proton decay, or possibly on indirect hints, such as relations 
among the gaugino masses.

Models of unification in warped space, on the other hand, predict light 
particles whose quantum numbers are characteristic of an underlying 
enlarged symmetry.  In the 5D description of these models, this enlarged 
symmetry is a grand unified gauge symmetry realized in the bulk, while in 
the dual 4D description, it is a global symmetry explicitly broken by the 
standard model gauge interactions.  A complete, realistic model of warped 
supersymmetric unification was constructed in Ref.~\cite{Goldberger:2002pc}, 
building on an earlier suggestion~\cite{Pomarol:2000hp}, and on subsequent 
developments in our understanding of gauge coupling evolution in 
warped space~[\ref{Randall:2001gc:X}~--~\ref{Contino:2002kc:X}] (a 
non-supersymmetric model has been discussed in~\cite{Agashe:2002pr}). 
As emphasized in~\cite{Goldberger:2002pc}, this framework leaves many 
of the most attractive features of conventional unification intact. 
For example, in the minimal supersymmetric standard model (MSSM), the 
assumption of unification yields an accurate prediction relating the 
values of the gauge couplings at low energies; in warped supersymmetric 
unification, a strong-coupling assumption leads to the same prediction 
at leading-log level.  Yet physics at accessible energies could be quite 
different than in the conventional scenario.  The model reveals its 
higher-dimensional nature at the TeV scale, through the appearance 
of Kaluza-Klein (KK) towers and an $N=2$ supermultiplet structure.

In this paper we construct a model of warped unification in which the
deviations from conventional expectations are perhaps even more striking 
than in~\cite{Goldberger:2002pc}.  Before supersymmetry breaking, the 
{\em massless} spectrum of the theory includes not only the states of 
the MSSM, but also exotic descendants of the enlarged symmetry.  As in 
the model of~\cite{Goldberger:2002pc}, however, these extra particles 
do not spoil gauge coupling unification.  From the 5D perspective, this 
setup arises when the grand unified gauge symmetry is broken by boundary 
conditions both on the Planck brane and on the TeV brane. From the 4D 
perspective, it arises when the approximate global symmetry is broken 
spontaneously by the strong dynamics associated with the conformal sector. 
The extra massless states make up the supermultiplets containing the 
pseudo-Goldstone bosons associated with this spontaneous breakdown.

We assume that the strong dynamics that break the global symmetry of 
the conformal sector also break supersymmetry, thereby giving masses 
to the particles in the pseudo-Goldstone multiplets.  This link between 
the spontaneous breakdown of supersymmetry and the spontaneous breakdown 
of the enlarged symmetry is another important difference between this 
model and more conventional supersymmetric unification.  Here there 
is no reason why the underlying unified symmetry should leave its imprint 
on the MSSM gaugino masses in any way --- the bino might even be heavier 
than the gluino.  From the 5D perspective,  this is simply a consequence 
of the fact that the various supersymmetry-breaking gaugino masses on 
the TeV brane are completely independent, as the unified symmetry is 
not realized there.  As a consequence, squark and slepton masses also 
do not obey the pattern characteristic to typical unified theories.  This 
is in contrast to the situation in the model of~\cite{Goldberger:2002pc}, 
where the masses of all superparticles are characterized essentially 
by a single unified supersymmetry-breaking gaugino mass 
term~\cite{Nomura:2003qb}. 

The model we will study, then, is a supersymmetric extension of the 
standard model that incorporates gauge coupling unification, has a 
massless spectrum in the supersymmetric limit that differs substantially 
from that of the MSSM (the extra states do not even fill out full $SU(5)$ 
multiplets), and  features a superpartner spectrum that generically looks 
nothing like those usually encountered in supersymmetric unification. 
We will see that such a model can be made fully realistic by introducing 
TeV-brane operators that transmit supersymmetry breaking to the 
pseudo-Goldstone multiplet.  We will also calculate the spectrum 
of light states, and identify certain characteristic features of 
the superparticle spectrum that may be tested at future collider 
experiments. 

The organization of the paper is as follows.  In section~\ref{sec:model} 
we construct our model.  Supersymmetry breaking and its effects on various 
multiplets are discussed in section~\ref{sec:susy-breaking}, and the 
spectrum is calculated.  The phenomenology of the model is discussed 
in section~\ref{sec:pheno}.  The model presented in this paper does not 
explicitly address the issues of charge quantization or quark-lepton 
unification.  A model accommodating these features will be presented 
in a separate paper~\cite{NS}.

\section{Model}
\label{sec:model}

In this section we construct our model.  The construction closely 
follows that of Ref.~\cite{Goldberger:2002pc}.  The model is formulated 
in a 5D warped spacetime with the extra dimension compactified on an 
$S^1/Z_2$ orbifold: $0 \leq y \leq \pi R$, where $y$ represents the 
coordinate of the extra dimension.  The metric is given by 
\begin{equation}
  d s^2 = e^{-2k|y|} \eta_{\mu\nu} dx^\mu dx^\nu + dy^2,
\label{eq:metric}
\end{equation}
where $k$ is the AdS curvature, which is taken to be somewhat (typically 
a factor of a few) smaller than the 5D Planck scale $M_5$.  The 4D Planck 
scale, $M_{\rm Pl}$, is given by $M_{\rm Pl}^2 \simeq M_5^3/k$ and we 
take $k \sim M_5 \sim M_{\rm Pl}$.  We choose $kR \sim 10$ so that the 
TeV scale is naturally generated by the AdS warp factor: $k' \equiv 
k e^{-\pi kR} \sim {\rm TeV}$~\cite{Randall:1999ee}.%
\footnote{The quantity $k'$ was denoted as $T$ in 
Refs.~\cite{Goldberger:2002pc,Nomura:2003qb}.}

We consider a supersymmetric $SU(5)$ gauge theory on the above 
gravitational background.  The bulk $SU(5)$ symmetry is broken by 
boundary conditions at both $y=0$ and $\pi R$.  Specifically, the 5D 
gauge multiplet can be decomposed into a 4D $N=1$ vector superfield 
$V(A_\mu, \lambda)$ and a 4D $N=1$ chiral superfield $\Sigma(\sigma+iA_5, 
\lambda')$, where both $V$ and $\Sigma$ are in the adjoint representation 
of $SU(5)$. The boundary conditions for these fields are given by
\begin{equation}
  \pmatrix{V \cr \Sigma}(x^\mu,-y) 
  = \pmatrix{P V P^{-1} \cr -P \Sigma P^{-1}}(x^\mu,y), 
\qquad
  \pmatrix{V \cr \Sigma}(x^\mu,-y') 
  = \pmatrix{P V P^{-1} \cr -P \Sigma P^{-1}}(x^\mu,y'), 
\label{eq:bc-g}
\end{equation}
where $y' = y - \pi R$, and $P$ is a $5 \times 5$ matrix acting on 
gauge space: $P = {\rm diag}(+,+,+,-,-)$.  This reduces the gauge 
symmetry to $SU(3)_C \times SU(2)_L \times U(1)_Y$ (321) both at 
the $y=0$ brane (Planck brane) and at the $y=\pi R$ brane (TeV brane). 
The gauge symmetry at low energies is $SU(3)_C \times SU(2)_L \times 
U(1)_Y$.  The zero-mode sector contains not only the 321 component of 
$V$, but also the $SU(5)/(SU(3)_C \times SU(2)_L \times U(1)_Y)$ (XY) 
component of $\Sigma$. The typical mass scale for the KK towers is 
$k' \sim {\rm TeV}$, so that the lowest KK excitations of the standard 
model gauge fields and the lightest XY gauge bosons both have masses 
of order TeV. 

The Higgs fields are introduced in the bulk as two hypermultiplets 
transforming as the fundamental representation of $SU(5)$.  Using 
notation where a hypermultiplet is represented by two 4D $N=1$ 
chiral superfields $\Phi(\phi,\psi)$ and $\Phi^c(\phi^c,\psi^c)$ with 
opposite gauge transformation properties, our two Higgs hypermultiplets 
can be written as $\{ H, H^c \}$ and $\{ \bar{H}, \bar{H}^c \}$, where 
$H$ and $\bar{H}^c$ transform as ${\bf 5}$ and $\bar{H}$ and $H^c$ 
transform as ${\bf 5}^*$ under $SU(5)$. The boundary conditions are 
given by 
\begin{equation}
  \pmatrix{H \cr H^c}(x^\mu,-y) 
  = \pmatrix{-P H \cr P H^c}(x^\mu,y), 
\qquad
  \pmatrix{H \cr H^c}(x^\mu,-y') 
  = \pmatrix{-P H \cr P H^c}(x^\mu,y'), 
\label{eq:bc-h}
\end{equation}
for $\{ H, H^c \}$, and similarly for $\{ \bar{H}, \bar{H}^c \}$. 
The zero modes consist of the $SU(2)_L$-doublet components of $H$ 
and $\bar{H}$ and the $SU(3)_C$-triplet components of $H^c$ and 
$\bar{H}^c$.  A bulk hypermultiplet $\{ \Phi, \Phi^c \}$ can 
generically have a mass term in the bulk, which is written as 
\begin{equation}
  S = \int\!d^4x \int_0^{\pi R}\!\!dy \, 
    \biggl[ e^{-3k|y|}\! \int\!d^2\theta\, c_\Phi k \Phi \Phi^c 
    + {\rm h.c.} \biggr],
\label{eq:bulk-mass}
\end{equation}
in the basis where the kinetic term is given by $S_{\rm kin} = \int\!d^4x 
\int\!dy\, [e^{-2k|y|} \int\!d^4\theta (\Phi^\dagger \Phi + \Phi^c 
\Phi^{c\dagger}) + \{ e^{-3k|y|} \int\!d^2\theta (\Phi^c \partial_y \Phi 
- \Phi \partial_y \Phi^c)/2 + {\rm h.c.} \}]$~\cite{Marti:2001iw}.  The 
parameter $c_\Phi$ controls the wavefunction profile of the zero mode. 
For $c_\Phi > 1/2$ ($< 1/2$) the wavefunction of a zero mode arising 
from $\Phi$ is localized to the Planck (TeV) brane; for $c_\Phi = 1/2$ 
it is conformally flat.  If a zero mode arises from $\Phi^c$, its 
wavefunction is localized to the TeV (Planck) brane for $c_\Phi > -1/2$ 
($< -1/2$) and conformally flat for $c_\Phi = -1/2$.  For the Higgs 
fields, we choose $c_H, c_{\bar{H}} \geq 1/2$ to preserve the MSSM 
prediction for gauge coupling unification (see below).%
\footnote{An alternative choice for the boundary conditions of the Higgs 
fields is given by Eq.~(\ref{eq:bc-h}) with an extra minus sign in 
the right-hand side of the second equation (i.e.~flipping the TeV-brane 
boundary conditions).  This also gives the MSSM prediction for gauge 
coupling unification for $c_H, c_{\bar{H}} \geq 1/2$, since the 
prediction does not depend on the physics at the TeV brane.  Although 
these boundary conditions do not give a zero mode, four doublet states 
from $H$, $H^c$, $\bar{H}$ and $\bar{H}^c$ are relatively light [they 
are even exponentially lighter than $k'$ for $c_H, c_{\bar{H}} > 1/2$ with 
the modes from $H$ and $\bar{H}$ ($H^c$ and $\bar{H}^c$) exponentially 
localized to the Planck (TeV) brane]. A realistic model is then obtained 
by giving a mass term of the form $\int\!d^2\theta H_D^c \bar{H}_D^c 
+ {\rm h.c.}$ on the TeV brane.}

Matter fields are introduced on the Planck brane as a standard 
set of chiral superfields $Q({\bf 3}, {\bf 2})_{1/6}$, $U({\bf 3}^*, 
{\bf 1})_{-2/3}$, $D({\bf 3}^*, {\bf 1})_{1/3}$, $L({\bf 1}, 
{\bf 2})_{-1/2}$ and $E({\bf 1}, {\bf 1})_{1}$ for each generation. 
Here the numbers represent the transformation properties under 
$SU(3)_C \times SU(2)_L \times U(1)_Y$ with the $U(1)_Y$ charges 
normalized in the conventional way.  With matter localized on the 
Planck brane, proton decay can be adequately suppressed despite the 
fact that XY gauge and colored Higgs fields have masses of order TeV. 
This is because the wavefunctions of the XY gauge and colored Higgs 
fields (and their KK excitations) are all strongly localized to the 
TeV brane.  Yukawa couplings are introduced on the Planck brane:%
\footnote{Our convention for the delta function is $\int_0^\epsilon 
\delta(y) dy = 1/2$ for $\epsilon > 0$.}
\begin{equation}
  S = \int\!d^4x \int_0^{\pi R}\!\!dy \,\, 
    2 \delta(y) \biggl[ \int\!d^2\theta \left( y_u Q U H_D 
    + y_d Q D \bar{H}_D + y_e L E \bar{H}_D \right)
    + {\rm h.c.} \biggr],
\label{eq:yukawa-321}
\end{equation}
where $H_D$ and $\bar{H}_D$ are the doublet components of $H$ and 
$\bar{H}$.  The Yukawa couplings respect a $U(1)_R$ symmetry, under 
which the 4D superfields $V, \Sigma, H$ and $\bar{H}$ are neutral, 
$Q, U, D, L$ and $E$ have unit charge, and $H^c$ and $\bar{H}^c$ have 
charge $+2$, and we impose this symmetry on the theory.  This $U(1)_R$ 
forbids dangerous dimension four and five proton decay operators 
together with a potentially large supersymmetric mass term for the 
Higgs fields~\cite{Hall:2001pg} (the $U(1)_R$ symmetry is broken to its 
$Z_2$ subgroup through supersymmetry breaking but without reintroducing 
phenomenological problems).  Small neutrino masses can be naturally 
generated through the conventional seesaw mechanism by introducing 
right-handed neutrino fields $N({\bf 1}, {\bf 1})_{0}$ with the Majorana 
mass terms and neutrino Yukawa couplings on the Planck brane:
\begin{equation}
  S = \int\!d^4x \int_0^{\pi R}\!\!dy \,\, 
    2 \delta(y) \biggl[ \int\!d^2\theta \left( \frac{M_N}{2} N N 
    + y_\nu L N H_D \right) + {\rm h.c.} \biggr],
\label{eq:neutrino}
\end{equation}
where $N$ carries a $U(1)_R$ charge of $+1$.

We note that in our theory there is a priori no reason why the $U(1)_Y$ 
charges for the matter fields must obey the usual $SU(5)$ normalization 
(the one for which the Yukawa couplings of Eq.~(\ref{eq:yukawa-321}) 
are allowed).  Thus our theory is not ``grand unified'' in the conventional 
sense.  The correct normalization may be obtained by considering higher 
dimensional theories as in~\cite{Hall:2002qw}; for instance by extending 
the $y=0$ brane (the bulk) to a short 1-dimensional (thin 2-dimensional) 
object having $SU(4)_C \times SU(2)_L \times SU(2)_R$ ($SO(10)$) gauge 
symmetry.  Alternatively, we could consider the conventional ``grand 
unified theory'' version of our theory, i.e. we could break $SU(5)$ by the 
brane-localized Higgs field at $y=0$, although we then need some mechanism 
for doublet-triplet splitting on the brane.  Finally, one could also 
consider a 5D theory with an enlarged bulk gauge group such as $SO(10)$, 
which will be discussed in Ref.~\cite{NS}.

Despite the presence of exotic low-energy states, the model still 
reproduces the successful prediction associated with MSSM gauge coupling 
unification.  From the 5D viewpoint, there are three local operators 
that can contribute to the low-energy 4D gauge couplings:
\begin{equation}
  S = -{1\over 4}\int\!d^4x \int_0^{\pi R}\!\!dy \, 
    \biggl[ \frac{1}{g_B^2} F_{\mu\nu} F^{\mu\nu} 
    + 2 \delta(y) \frac{1}{\tilde{g}_{0,a}^2} {F^a}_{\mu\nu} {F^a}^{\mu\nu} 
    + 2 \delta(y - \pi R) \frac{1}{\tilde{g}_{\pi,a}^2} 
    {F^a}_{\mu\nu} {F^a}^{\mu\nu} \biggr],
\label{eq:gen-kin}
\end{equation}
where $g_B$ is the $SU(5)$-invariant 5D gauge coupling and the index 
$a$ runs over $SU(3)_C$, $SU(2)_L$ and $U(1)_Y$ ($a=3,2,1$, respectively). 
The structure of these terms is determined by the restricted 5D gauge 
symmetry, which reduces to 321 on the $y=0$ and $y=\pi R$ branes but 
is $SU(5)$ in the bulk.  At the fundamental scale $M_* \sim M_5$, the 
coefficients of these operators are incalculable parameters of the 
effective field theory, so that one might worry that the model may not 
give any prediction for the low-energy gauge couplings. This difficulty, 
however, can be avoided if we require that the entire theory is strongly 
coupled at the scale $M_*$.  In this case the sizes of these coefficients 
are estimated as $1/g_B^2 \simeq M_*/16\pi^3$ and $1/\tilde{g}_{0,a}^2 
\simeq 1/\tilde{g}_{\pi,a}^2 \simeq 1/16\pi^2$, and one finds that the 
low-energy prediction is insensitive to the parameters $\tilde{g}_{0,a}$ 
and $\tilde{g}_{\pi,a}$ evaluated at $M_*$.  The prediction for the 
low-energy 4D gauge couplings, $g_a$, is then written in the form
\begin{equation}
  \frac{1}{g_a^2(k')} 
  \simeq (SU(5)\,\,\, {\rm symmetric}) 
    + \frac{1}{8 \pi^2} \Delta^a(k',k),
\label{eq:gc-low-1}
\end{equation}
where $\Delta^a(k',k)$ is the quantity whose non-universal part can be 
unambiguously computed in the effective theory.  In the present model, 
this quantity is given at one-loop leading-log level by setting 
$(T_1,T_2,T_3)(V_{++}) = (0,2,3)$, $(T_1,T_2,T_3)(V_{--}) = (5,3,2)$, 
$(T_1,T_2,T_3)(V_{+-}) = (T_1,T_2,T_3)(V_{-+}) = (0,0,0)$ and 
$(T_1,T_2,T_3)(\Phi_{++}) = (3/10,1/2,0)$, $(T_1,T_2,T_3)(\Phi_{--}) 
= (1/5,0,1/2)$, $(T_1,T_2,T_3)(\Phi_{+-}) = (T_1,T_2,T_3)(\Phi_{-+}) 
= (0,0,0)$, $c_{++} = c_{--} \geq 1/2$ for $\Phi = H, \bar{H}$ in 
Eqs.~(9) and (10) of Ref.~\cite{Goldberger:2002pc}, respectively. 
Adding everything together, we obtain
\begin{equation}
  \pmatrix{\Delta^1 \cr \Delta^2 \cr \Delta^3}(k', k)
    \simeq \pmatrix{33/5 \cr 1 \cr -3} \ln\left(\frac{k}{k'}\right),
\label{eq:gc-low-2}
\end{equation}
where we have absorbed a possible $SU(5)$ symmetric piece into the first 
term of Eq.~(\ref{eq:gc-low-1}).  This is exactly the relation obtained 
in conventional 4D supersymmetric unification with the parameter $k$ 
identified with the unification scale.  This result can be understood 
more intuitively by noticing that the gauge couplings above the TeV scale 
can be defined as the coefficients of the gauge two-point correlators 
whose end points are both on the Planck brane~\cite{Goldberger:2002cz}. 
We then find that the light extra states do not contribute to the large 
logarithm in Eq.~(\ref{eq:gc-low-2}) because they are all strongly 
localized to the TeV brane and the Planck-brane gauge correlators do 
not probe the region near the TeV brane at energies higher than $k'$. 
In a suitable renormalization scheme, the large logarithmic contribution 
can be absorbed in the couplings on the Planck brane.  In this case the 
brane couplings renormalized at the scale $\mu \sim k'$ are given by
\begin{equation}
  \frac{1}{\tilde{g}_{0,a}^2(k')} 
    \simeq \frac{1}{8\pi^2}\Delta^a(k',k), 
\qquad
  \frac{1}{\tilde{g}_{\pi,a}^2(k')} 
    = O\Bigl(\frac{1}{16\pi^2}\Bigr),
\label{eq:gc-low-3}
\end{equation}
where $\Delta^a(k',k)$ are given by Eq.~(\ref{eq:gc-low-2}) and the scales 
are measured in terms of the 4D metric $\eta_{\mu\nu}$~\cite{Nomura:2003qb}. 
We can thus safely neglect $1/\tilde{g}_{\pi,a}^2$ in any formulae given 
below (and we will) but not necessarily $1/\tilde{g}_{0,a}^2$. 

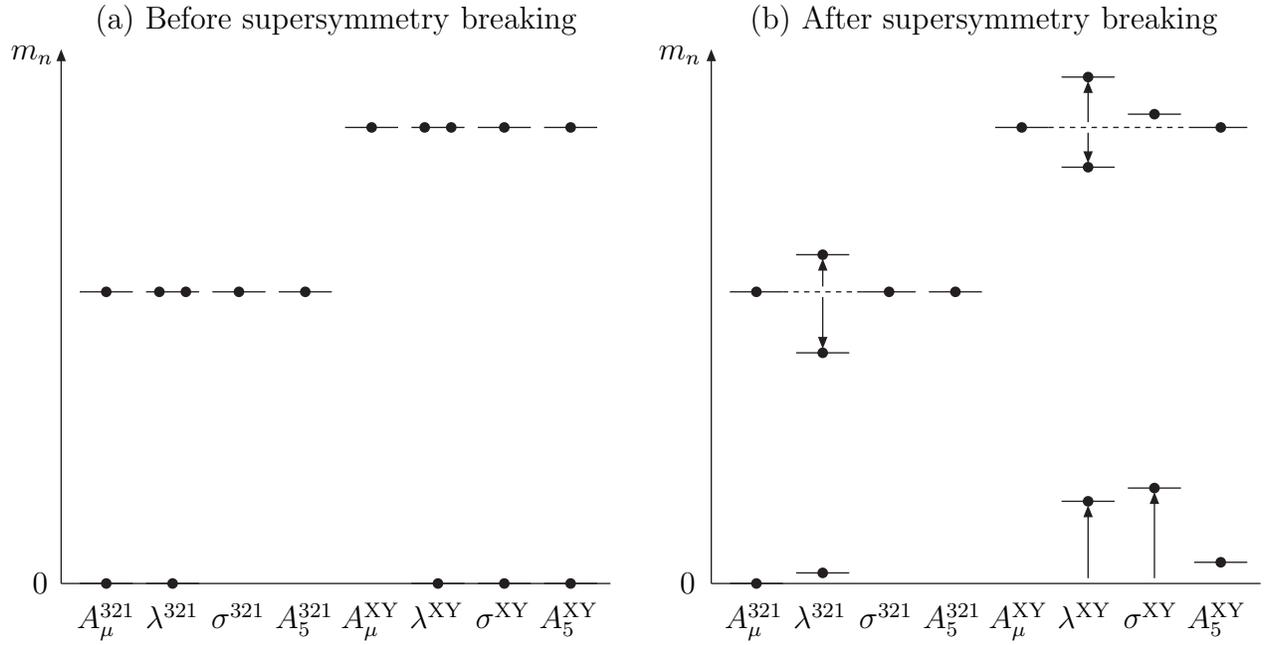
\begin{figure}
\begin{center}
\begin{picture}(470,240)(-20,-20)
 \Text(102,210)[b]{(a) Before supersymmetry breaking}
 \Line(-2,0)(205,0)
 \LongArrow(-2,0)(-2,200) \Text(-5,200)[r]{$m_n$} \Text(-7,0)[r]{0}
 \Text(15,-17)[b]{$A_\mu^{321}$}
   \Line(5,0)(25,0) \Vertex(15,0){2}
   \Line(5,110)(25,110) \Vertex(15,110){2}
 \Text(40,-17)[b]{$\lambda^{321}$}
   \Line(30,0)(50,0) \Vertex(40,0){2}
   \Line(30,110)(50,110) \Vertex(35,110){2} \Vertex(45,110){2}
 \Text(65,-17)[b]{$\sigma^{321}$}
   \Line(55,110)(75,110) \Vertex(65,110){2}
 \Text(90,-17)[b]{$A_5^{321}$}
   \Line(80,110)(100,110) \Vertex(90,110){2}
 \Text(115,-17)[b]{$A_\mu^{\rm XY}$}
   \Line(105,172)(125,172) \Vertex(115,172){2}
 \Text(140,-17)[b]{$\lambda^{\rm XY}$}
   \Line(130,0)(150,0) \Vertex(140,0){2}
   \Line(130,172)(150,172) \Vertex(135,172){2} \Vertex(145,172){2}
 \Text(165,-17)[b]{$\sigma^{\rm XY}$}
   \Line(155,0)(175,0) \Vertex(165,0){2}
   \Line(155,172)(175,172) \Vertex(165,172){2}
 \Text(190,-17)[b]{$A_5^{\rm XY}$}
   \Line(180,0)(200,0) \Vertex(190,0){2}
   \Line(180,172)(200,172) \Vertex(190,172){2}
 \Text(347,210)[b]{(b) After supersymmetry breaking}
 \Line(243,0)(450,0)
 \LongArrow(243,0)(243,200) \Text(240,200)[r]{$m_n$} \Text(238,0)[r]{0}
 \Text(260,-17)[b]{$A_\mu^{321}$}
   \Line(250,0)(270,0) \Vertex(260,0){2}
   \Line(250,110)(270,110) \Vertex(260,110){2}
 \Text(285,-17)[b]{$\lambda^{321}$}
   \Line(275,4)(295,4) \Vertex(285,4){2}
   \Line(275,87)(295,87) \Vertex(285,87){2} \LongArrow(285,108)(285,90)
   \Line(275,124)(295,124) \Vertex(285,124){2} \LongArrow(285,112)(285,121)
 \Text(310,-17)[b]{$\sigma^{321}$}
   \Line(300,110)(320,110) \Vertex(310,110){2}
 \Text(335,-17)[b]{$A_5^{321}$}
   \Line(325,110)(345,110) \Vertex(335,110){2}
 \DashLine(270,110)(300,110){2}
 \Text(360,-17)[b]{$A_\mu^{\rm XY}$}
   \Line(350,172)(370,172) \Vertex(360,172){2}
 \Text(385,-17)[b]{$\lambda^{\rm XY}$}
   \Line(375,31)(395,31) \Vertex(385,31){2} \LongArrow(385,2)(385,28)
   \Line(375,157)(395,157) \Vertex(385,157){2} \LongArrow(385,170)(385,160)
   \Line(375,191)(395,191) \Vertex(385,191){2} \LongArrow(385,174)(385,188)
 \Text(410,-17)[b]{$\sigma^{\rm XY}$}
   \Line(400,36)(420,36) \Vertex(410,36){2} \LongArrow(410,2)(410,33)
   \Line(400,177)(420,177) \Vertex(410,177){2}
 \Text(435,-17)[b]{$A_5^{\rm XY}$}
   \Line(425,8)(445,8) \Vertex(435,8){2}
   \Line(425,172)(445,172) \Vertex(435,172){2}
 \DashLine(370,172)(425,172){2}
\end{picture}
\caption{Schematic depiction for the lowest-lying masses for the 321 gauge 
 multiplet ($A_\mu^{321}$, $\lambda^{321}$, $\sigma^{321}$, $A_5^{321}$) 
 and the XY gauge multiplet ($A_\mu^{\rm XY}$, $\lambda^{\rm XY}$, 
 $\sigma^{\rm XY}$, $A_5^{\rm XY}$) (a) before and (b) after supersymmetry 
 breaking.  Each bullet for $\lambda^{321}$ and $\lambda^{\rm XY}$ 
 represents a Majorana and Dirac degree of freedom, respectively.  Arrows 
 indicate displacements of the mass levels relative to their supersymmetric 
 positions, represented by dotted lines.}
\label{fig:spectrum}
\end{center}
\end{figure}

Let us now work out the spectrum of the model in more detail, following 
the procedure of Ref.~\cite{Nomura:2003qb}.  We first consider the 
gauge sector.  The zero modes consist of the 321 component of $V$, 
$V^{321}$, and the XY component of $\Sigma$, $\Sigma^{\rm XY}$, which 
transform as $({\bf 8}, {\bf 1})_0 + ({\bf 1}, {\bf 3})_0 + ({\bf 1}, 
{\bf 1})_0$ and $({\bf 3}, {\bf 2})_{-5/6} + ({\bf 3}^*, {\bf 2})_{5/6}$ 
under $SU(3)_C \times SU(2)_L \times U(1)_Y$, respectively.  The KK 
excitations for the 321 fields consist of $V^{321}$ and $\Sigma^{321}$ 
at each KK level, whose masses $m_n$ are determined by the equation 
\begin{equation}
  \frac{J_0\left(\frac{m_n}{k}\right)
    + \frac{g_B^2}{\tilde{g}_{0,a}^2}m_n J_1\left(\frac{m_n}{k}\right)}
  {Y_0\left(\frac{m_n}{k}\right)
    + \frac{g_B^2}{\tilde{g}_{0,a}^2}m_n Y_1\left(\frac{m_n}{k}\right)}
  = \frac{J_0\left(\frac{m_n}{k'}\right)}
    {Y_0\left(\frac{m_n}{k'}\right)},
\label{eq:KKmass-321}
\end{equation}
where $J_n(x)$ and $Y_n(x)$ are the Bessel functions of order $n$, and 
$\tilde{g}_{0,a}^2$ ($a=1,2,3$) are given by Eq.~(\ref{eq:gc-low-3}).%
\footnote{The effects of brane-localized kinetic terms on the spectrum 
of the gauge boson KK towers were studied in~\cite{Carena:2002dz}.}
For the XY fields, each KK excited level consists of $V^{\rm XY}$ and 
$\Sigma^{\rm XY}$ with the masses given by
\begin{equation}
  \frac{J_1\left(\frac{m_n}{k}\right)}
    {Y_1\left(\frac{m_n}{k}\right)}
  = \frac{J_1\left(\frac{m_n}{k'}\right)}
    {Y_1\left(\frac{m_n}{k'}\right)}.
\label{eq:KKmass-XY}
\end{equation}
Therefore, for $k' \ll k$, the spectrum in the gauge sector is given by
\begin{equation}
  \left\{ \begin{array}{ll} 
    V^{321}: & m_0 = 0, \\
    \{ V^{321}, \Sigma^{321} \}: 
      & m_n \simeq (n-\frac{1}{4})\pi k',
  \end{array} \right.
\qquad
  \left\{ \begin{array}{ll} 
    \Sigma^{\rm XY}: & m_0 = 0, \\
    \{ V^{\rm XY}, \Sigma^{\rm XY} \}: 
      & m_n \simeq (n+\frac{1}{4})\pi k',
  \end{array} \right.
\label{eq:spectrum-gauge}
\end{equation}
where $n = 1,2,\cdots$.  This spectrum is depicted for the lowest-lying 
modes in Fig.~\ref{fig:spectrum}a.  The resulting spectrum is quite 
different from that of the model discussed in~\cite{Goldberger:2002pc,%
Nomura:2003qb}.  The massless sector contains the XY states $\Sigma^{\rm XY}$ 
as well as the MSSM gauge fields $V^{321}$, and the KK excited states 
are not $SU(5)$ symmetric even approximately.  As we will see in the next 
section, the unwanted zero modes from $\Sigma^{\rm XY}$ obtain masses 
once supersymmetry is broken. 

For the Higgs fields, the massless sector consists of two doublets $H_D$ 
and $\bar{H}_D$ ($\subset H$ and $\bar{H}$) and two triplets $H_T^c$ and 
$\bar{H}_T^c$ ($\subset H^c$ and $\bar{H}^c$).  The excited states for the 
doublets and triplets contain $\{ H_D, \bar{H}_D, H_D^c, \bar{H}_D^c \}$ 
and $\{ H_T, \bar{H}_T, H_T^c, \bar{H}_T^c \}$, respectively.  For 
$\{ H, H^c \}$, the KK masses are given by 
\begin{equation}
  \frac{J_{c_H \mp 1/2}\left(\frac{m_n}{k}\right)}
    {Y_{c_H \mp 1/2}\left(\frac{m_n}{k}\right)}
  = \frac{J_{c_H \mp 1/2}\left(\frac{m_n}{k'}\right)}
    {Y_{c_H \mp 1/2}\left(\frac{m_n}{k'}\right)},
\label{eq:KKmass-Higgs}
\end{equation}
where $\mp$ takes $-$ and $+$ for the doublets and triplets, respectively. 
Similarly, the masses for $\{ \bar{H}, \bar{H}^c \}$ are given by 
Eq.~(\ref{eq:KKmass-Higgs}) with $c_H$ replaced by $c_{\bar{H}}$.  Here, 
we have neglected possible Planck-brane localized kinetic terms for 
simplicity.  The Higgs spectrum is thus given by 
\begin{equation}
  \left\{ \begin{array}{ll} 
    \{ H_D, \bar{H}_D \}: & m_0 = 0, \\
    \{ H_D, H_D^c \}: 
      & m_n \simeq (n+\frac{c_H}{2}-\frac{1}{2})\pi k', \\
    \{ \bar{H}_D, \bar{H}_D^c \}: 
      & m_n \simeq (n+\frac{c_{\bar{H}}}{2}-\frac{1}{2})\pi k',
  \end{array} \right.
\qquad
  \left\{ \begin{array}{ll} 
    \{ H_T^c, \bar{H}_T^c \}: & m_0 = 0, \\
    \{ H_T, H_T^c \}: 
      & m_n \simeq (n+\frac{c_H}{2})\pi k', \\
    \{ \bar{H}_T, \bar{H}_T^c \}: 
      & m_n \simeq (n+\frac{c_{\bar{H}}}{2})\pi k',
  \end{array} \right.
\label{eq:spectrum-Higgs}
\end{equation}
where $n = 1,2,\cdots$.  For $c_H = c_{\bar{H}} = 1/2$ the Higgs spectrum 
is identical to the gauge spectrum with the doublet and triplet components 
corresponding to the 321 and XY components, respectively (up to small 
difference arising from the Planck-brane localized operators).  For 
$c_H, c_{\bar{H}} > 1/2$, the Higgs KK towers are heavier than the gauge 
towers.  After supersymmetry breaking, the $\{ H, H^c \}$ and $\{ \bar{H}, 
\bar{H}^c \}$ states are mixed and the spectrum is distorted through 
the supersymmetric and supersymmetry-breaking mass terms generated on the 
TeV brane (or on the Planck brane depending on the details of the Higgs 
sector).  The zero modes in Eq.~(\ref{eq:spectrum-Higgs}) obtain masses 
through these operators. 

It is useful to understand the structure of the model described 
above in terms of the 4D dual description of the theory.  Through 
the AdS/CFT duality~\cite{Maldacena:1997re}, applied to a truncated 
space~\cite{Arkani-Hamed:2000ds}, we can relate our 5D model to a purely 
4D theory.  The 4D theory is defined at the ultraviolet (UV) cutoff scale 
of order $k \sim M_{\rm Pl}$ and contains a gauge interaction with the 
group $G$, whose coupling $\tilde{g}$ evolves very slowly over a wide 
energy interval below $k$.  Denoting the size of the group $G$ to be $N$, 
the correspondence is given by $\tilde{g}^2 N/16\pi^2 \approx M_*/\pi k$ 
and $N \approx 16\pi^2/g_B^2 k$ (so $\tilde{g} \simeq 4\pi$ and $N 
\simgt 1$ here).  The bulk gauge symmetry and the Planck-brane boundary 
conditions in the 5D theory imply that the $G$ gauge sector possesses 
a global $SU(5)$ symmetry whose $SU(3)_C \times SU(2)_L \times U(1)_Y$ 
subgroup is explicitly gauged.  The fields singlet under $G$ correspond 
to the modes localized to the Planck brane, i.e.~the MSSM quark, lepton 
and Higgs fields.  The theory below $k$, therefore, appears as a 
supersymmetric $SU(3)_C \times SU(2)_L \times U(1)_Y \times G$ gauge 
theory with the quarks, leptons and two Higgs doublets transforming 
under $SU(3)_C \times SU(2)_L \times U(1)_Y$.  In this 4D picture, 
a prediction relating the gauge couplings at low energies arises because 
the gauge couplings of $SU(3)_C$, $SU(2)_L$ and $U(1)_Y$ become strong 
at the UV scale $k$ due to the asymptotically non-free contribution from 
$G$~\cite{Goldberger:2002pc}.  Since the $G$ contribution is $SU(5)$ 
symmetric, the prediction is determined by the contributions from the 
matter, Higgs and 321 gauge multiplets, reproducing the successful 
MSSM prediction. 

At the TeV scale the gauge interaction of $G$ becomes strong and exhibits 
non-trivial infrared (IR) phenomena, corresponding to the presence of the 
TeV brane in the 5D picture.  The boundary conditions of Eq.~(\ref{eq:bc-g}) 
then imply that the global $SU(5)$ symmetry of the $G$ sector is spontaneously 
broken to $SU(3)_C \times SU(2)_L \times U(1)_Y$ by the IR dynamics of $G$. 
The IR dynamics of $G$ also produces resonances of masses of order TeV. 
These resonances have a tower structure and correspond to the KK towers 
in the 5D picture.  Because the global $SU(5)$ of $G$ is dynamically 
broken, the spectrum of the towers does not respect $SU(5)$, as shown in 
Eqs.~(\ref{eq:spectrum-gauge},~\ref{eq:spectrum-Higgs}).  Moreover, 
the dynamical breaking of $SU(5)$ to 321 produces Goldstone bosons, 
which correspond to the XY components of $A_5$ in an appropriate basis. 
This implies that in the absence of supersymmetry breaking the entire 
4D supermultiplet containing $A_5^{\rm XY}$ (i.e.~$\Sigma^{\rm XY}$) must 
be massless, in agreement with Eq.~(\ref{eq:spectrum-gauge}).  In fact, 
because of the partial gauging of $SU(5)$, the global $SU(5)$ of $G$ is 
explicitly broken by the 321 gauge interactions and the $A_5^{\rm XY}$ 
is only a pseudo-Goldstone boson.  This, however, does not give any mass 
for $\Sigma^{\rm XY}$ as long as supersymmetry is unbroken.

It is now relatively clear what happens when supersymmetry is broken. 
Supersymmetry breaking can be caused by the dynamics of $G$, which is
the situation we will consider in the next section.  It then gives masses 
of order TeV to $\sigma^{\rm XY}$ and $\lambda'^{\rm XY}$, that is, to 
all of $\Sigma^{\rm XY}$ except for $A_5^{\rm XY}$.  The breaking of 
supersymmetry is accompanied by that of $U(1)_R$, and the MSSM gauginos 
and Higgs multiplets also obtain masses.  Squarks and sleptons are massless 
at the leading order because they do not directly interact with the $G$ 
sector.  Their masses, however, are generated at one loop through the 
321 gauge interactions.  Since these masses are flavor universal, the 
supersymmetric flavor problem is absent.  The mass of $A_5^{\rm XY}$ is 
also generated through the 321 gauge loop, since it picks up the explicit 
breaking of the global $SU(5)$. The only massless fields remaining after 
supersymmetry breaking are the standard-model quarks, leptons and gauge 
fields.  Both the MSSM superparticles and exotic grand-unified-theoretic 
(GUT) states such as $\Sigma^{\rm XY}$ obtain masses, although the exotic 
states are generically expected to be heavier because they are composite 
states of $G$ and thus more strongly coupled to supersymmetry breaking 
caused by $G$. 

In the next section we discuss supersymmetry breaking in more detail 
using the 5D picture.  It will be shown how the dynamical supersymmetry 
breaking scenario outlined above is realized in 5D. The spectrum of 
the lowest-lying modes and its various interesting features will also 
be worked out.

\section{Supersymmetry Breaking}
\label{sec:susy-breaking}

As we have seen in the previous section, the massless sector of our 
model before supersymmetry breaking contains a chiral superfield with 
the quantum numbers of the XY gauge bosons, $\Sigma^{\rm XY}$, and a 
vector-like pair of color-triplet Higgs fields, $H_T^c$ and $\bar{H}_T^c$, 
in addition to the usual MSSM fields.  After supersymmetry is broken, 
all these fields must obtain masses, along with the MSSM superparticles. 
In this section we discuss how these masses arise and what the spectrum 
of the extra states and the MSSM superparticles looks like.  We also 
present a complete set of formulae giving the masses for the light 
states as well as the KK towers. 

We consider the case where supersymmetry is broken on the TeV brane. 
In terms of the 4D picture, this corresponds to a situation in which 
supersymmetry is broken by the dynamics of $G$ at the TeV scale. Assuming 
that $U(1)_R$ is also broken by these dynamics, we can parameterize 
their effects by a supersymmetry (and $U(1)_R$-symmetry) breaking vacuum 
expectation value.  Specifically, we introduce a singlet chiral field $Z$ 
on the TeV brane together with the superpotential~\cite{Gherghetta:2000qt}:
\begin{equation}
  S = \int\!d^4x \int_0^{\pi R}\!\!dy \,\, 
    2 \delta(y-\pi R) \biggl[ e^{-2\pi kR}\! \int\!d^4\theta Z^\dagger Z
    + \biggl\{ e^{-3\pi kR}\! \int\!d^2\theta \Lambda^2 Z + 
      {\rm h.c.} \biggr\} \biggr],
\label{eq:Z-TeV}
\end{equation}
which produces the desired vacuum expectation value $\langle Z \rangle 
= - e^{-\pi kR} \Lambda^{*2} \theta^2$, where $\Lambda$ is a mass parameter 
of order $M_* \sim M_5$.  Here, we have assumed that higher powers in $Z$ 
are absent in the superpotential, and that the flat direction of $Z$ is 
stabilized by higher order terms in the K\"ahler potential (for a dynamical 
model achieving this, see e.g.~\cite{Izawa:1997gs}).  This will give 
TeV-scale masses for various extra states and the MSSM superparticles, 
which we will discuss in turn.  As was discussed in~\cite{Goldberger:2002pc}, 
this breaking does not disturb the prediction relating the low-energy 
gauge couplings.

Let us begin by the gauge sector of the model.  The gauge kinetic terms 
are given by~\cite{Marti:2001iw}%
\footnote{In this paper we give all the expressions in the 
Wess-Zumino gauge.}
\begin{eqnarray}
  S &=& \int\!d^4x \int_0^{\pi R}\!\!dy\,
    \Biggl[ \biggl\{ \frac{1}{2 g_B^2} \int\!d^2\theta\, 
      {\rm Tr}[ {\cal W}^\alpha {\cal W}_\alpha ] + {\rm h.c.} \biggr\}
    + \frac{e^{-2k|y|}}{2 g_B^2} \int\!d^4\theta\, {\rm Tr}[ {\cal A}^2 ] 
\nonumber \\
    && \qquad \qquad  \quad
    + 2 \delta(y) \!\!\sum_{a=1,2,3} \biggl\{ 
      \frac{1}{2 \tilde{g}_{0,a}^2} \int\!d^2\theta\, 
      {\rm Tr}[ {\cal W}_a^\alpha {\cal W}_{a \alpha} ] 
      + {\rm h.c.} \biggr\} \Biggr],
\label{eq:gauge-kin}
\end{eqnarray}
where
\begin{equation}
  {\cal A} \equiv e^{-V}\!(\partial_y e^V) + (\partial_y e^V\!)\,e^{-V}
    - \sqrt{2}\, e^V \Sigma\, e^{-V} - \sqrt{2}\, e^{-V} \Sigma^\dagger e^V,
\label{eq:def-A}
\end{equation}
${\cal W}_\alpha \equiv -(1/8)\bar{\cal D}^2(e^{-2V} {\cal D}_\alpha e^{2V})$ 
is the $SU(5)$ field-strength superfield, and ${\cal W}_{a \alpha}$ 
with $a=1,2,3$ are the field-strength superfields for the $U(1)_Y$, 
$SU(2)_L$ and $SU(3)_C$ subgroups, respectively (${\cal W}_{a \alpha} 
\subset {\cal W}_\alpha$).  The MSSM gaugino masses then arise from 
the following operators:
\begin{equation}
  S = \int\!d^4x \int_0^{\pi R}\!\!dy\,
    2 \delta(y-\pi R) \!\!\sum_{a=1,2,3} \biggl[ 
      -\int\!d^2\theta\, \frac{\zeta_a}{M_*} Z \, 
      {\rm Tr}[ {\cal W}_a^\alpha {\cal W}_{a \alpha} ] 
      + {\rm h.c.} \biggr],
\label{eq:gaugino-mass}
\end{equation}
where $\zeta_a$ are dimensionless parameters.  Note that the operators 
in Eq.~(\ref{eq:gaugino-mass}) do not respect the $SU(5)$ symmetry, as 
$SU(5)$ is broken to 321 on the TeV brane by boundary conditions, and so 
these operators generate non-universal gaugino masses at the TeV scale.% 
\footnote{The mechanism of generating non-universal gaugino masses 
on an $SU(5)$-violating brane was considered in flat-space models 
in~\cite{Hall:2001pg}.}
Specifically, the masses of the 321 gauginos (and their KK towers) 
are given as the solution to the following equation:
\begin{equation}
  \frac{J_0\left(\frac{m_n}{k}\right)
    + \frac{g_B^2}{\tilde{g}_{0,a}^2}m_n J_1\left(\frac{m_n}{k}\right)}
  {Y_0\left(\frac{m_n}{k}\right)
    + \frac{g_B^2}{\tilde{g}_{0,a}^2}m_n Y_1\left(\frac{m_n}{k}\right)}
  = \frac{J_0\left(\frac{m_n}{k'}\right)
    + g_B^2 M_{\lambda,a} J_1\left(\frac{m_n}{k'}\right)}
    {Y_0\left(\frac{m_n}{k'}\right)
    + g_B^2 M_{\lambda,a} Y_1\left(\frac{m_n}{k'}\right)},
\label{eq:KKmass-321gauginos}
\end{equation}
where $M_{\lambda,a} \equiv \zeta_a \Lambda^{*2}/M_*$~\cite{Nomura:2003qb} 
(see also~\cite{Chacko:2003tf}).  Note that here we have to look for the 
solutions with $m_n < 0$ as well as those with $m_n > 0$ to obtain all the 
masses (the physical masses are given by $|m_n|$).  The non-universal nature 
of the gaugino masses becomes important when we discuss squark and slepton 
masses and the phenomenology of the model. 

The masses for fields in $\Sigma^{\rm XY}$ are generated through the 
operator
\begin{equation}
  S = \int\!d^4x \int_0^{\pi R}\!\!dy\,
    2 \delta(y-\pi R) \biggl[ e^{-2\pi kR} \int\!d^4\theta\, 
    \frac{\eta}{2 M_*} Z^\dagger \,
    {\rm Tr}[ {\cal P}[{\cal A}] {\cal P}[{\cal A}]] + {\rm h.c.} \biggr],
\label{eq:XYfermion-mass}
\end{equation}
where the trace is over the $SU(5)$ space and ${\cal P}[{\cal X}]$ 
is a projection operator: with ${\cal X}$ an adjoint of $SU(5)$, 
${\cal P}[{\cal X}]$ extracts the $({\bf 3}, {\bf 2})_{-5/6} + 
({\bf 3}^*, {\bf 2})_{5/6}$ component of ${\cal X}$ under the decomposition 
to 321. The coefficient $\eta$ is a dimensionless parameter.  Note that 
${\cal P}[{\cal A}]$ is even under the parity $y' \rightarrow -y'$ so 
that the above operator is non-vanishing. This operator gives masses 
to the fermion component $\lambda'^{\rm XY}$ and to the real-scalar 
component $\sigma^{\rm XY}$ contained in  $\Sigma^{\rm XY}$. The mass 
of $\lambda'^{\rm XY}$, and in fact the masses of the entire XY gaugino 
KK tower consisting of $\lambda^{\rm XY}$ and $\lambda'^{\rm XY}$, are 
given by the equation
\begin{equation}
  \frac{J_1\left(\frac{m_n}{k}\right)}
  {Y_1\left(\frac{m_n}{k}\right)}
  = \frac{J_1\left(\frac{m_n}{k'}\right)
    - g_B^2 M_{\lambda,X} J_0\left(\frac{m_n}{k'}\right)}
    {Y_1\left(\frac{m_n}{k'}\right)
    - g_B^2 M_{\lambda,X} Y_0\left(\frac{m_n}{k'}\right)},
\label{eq:KKmass-XYfermion}
\end{equation}
where $M_{\lambda,X} \equiv \eta \Lambda^2/M_*$.  Here, we have 
not included a possible kinetic term for $\Sigma^{\rm XY}$ on the 
Planck brane, but this term barely affects the spectrum since the 
$\lambda'^{\rm XY}$'s are strongly localized to the TeV brane.  For 
$\sigma^{\rm XY}$, we find a term proportional to $\delta(y-\pi R)^2$ 
in its equation of motion after integrating out the auxiliary field 
in $\Sigma^{\rm XY}$. This singular term, however, is canceled by 
appropriately choosing the coefficient of the operator 
\begin{equation}
  S = \int\!d^4x \int_0^{\pi R}\!\!dy\,
    2 \delta(y-\pi R) \biggl[ -e^{-2\pi kR} \int\!d^4\theta\, 
    \frac{\rho}{4 M_*^2} Z^\dagger Z \,
    {\rm Tr}[ {\cal P}[{\cal A}] {\cal P}[{\cal A}]] \biggr],
\label{eq:XYscalar-mass}
\end{equation}
which also gives a mass for $\sigma^{\rm XY}$. The consistency of the 
effective theory then requires $\rho$ to take the form $\rho = -8 g_B^2 
|\eta|^2 \delta(0) + \rho'$, where $\rho'$ is a dimensionless parameter. 
Here the first term in $\rho$ is chosen such that it cancels the singular 
term arising from the operator in Eq.~(\ref{eq:XYfermion-mass}).  In fact, 
this singular term in $\rho$ is simply a counterterm chosen to absorb 
divergences order by order in perturbation theory, although it appears 
already at tree level.  The masses of $\sigma^{\rm XY}$ and their KK 
towers are then given by 
\begin{equation}
  \frac{J_1\left(\frac{m_n}{k}\right)}
  {Y_1\left(\frac{m_n}{k}\right)}
  = \frac{J_1\left(\frac{m_n}{k'}\right)
    - \frac{g_B^2 M_{\sigma,X}^2 k'}{m_n k} J_0\left(\frac{m_n}{k'}\right)}
    {Y_1\left(\frac{m_n}{k'}\right)
    - \frac{g_B^2 M_{\sigma,X}^2 k'}{m_n k} Y_0\left(\frac{m_n}{k'}\right)},
\label{eq:KKmass-XYscalar}
\end{equation}
where $M_{\sigma,X}^2 \equiv \rho' |\Lambda|^4/M_*^2$.  The mass 
of $A_5^{\rm XY}$ does not directly arise from the operators in 
Eqs.~(\ref{eq:XYfermion-mass},~\ref{eq:XYscalar-mass}), as it is 
protected by the 5D gauge invariance (in the 4D picture $A_5^{\rm XY}$ 
is a pseudo-Goldstone boson of $SU(5) \rightarrow {\rm 321}$ caused by 
the dynamics of $G$).  However, it arises through loop effects as we 
will discuss later. 

The Higgs multiplets, $H_D$ and $\bar{H}_D$, obtain masses from the terms
\begin{eqnarray}
  S &=& \int\!d^4x \int_0^{\pi R}\!\!dy\,
    2 \delta(y-\pi R)\, e^{-2\pi kR} \int\!d^4\theta\, 
    \biggl[ \frac{\zeta_D^{}}{M_*^2} Z^\dagger H_D \bar{H}_D 
      + \frac{\zeta_D^*}{M_*^2} Z H_D^\dagger \bar{H}_D^\dagger 
\nonumber\\
  && - \frac{\rho_D^{}}{M_*^3} Z^\dagger Z H_D \bar{H}_D 
     - \frac{\rho_D^*}{M_*^3} Z^\dagger Z H_D^\dagger \bar{H}_D^\dagger 
     - \frac{\eta_{H_D}^{}}{M_*^3} Z^\dagger Z H_D^\dagger H_D
     - \frac{\eta_{\bar{H}_D}^{}}{M_*^3} Z^\dagger Z 
       \bar{H}_D^\dagger \bar{H}_D \biggr],
\label{eq:HD-mass}
\end{eqnarray}
where $\zeta_D^{}$, $\rho_D^{}$, $\eta_{H_D}^{}$ and $\eta_{\bar{H}_D}^{}$ 
are dimensionless parameters.  Similarly, the $H_T^c$ and $\bar{H}_T^c$ 
fields obtain masses from terms as in Eq.~(\ref{eq:HD-mass}) with 
the fields $\{H_D, \bar{H}_D\}$ and couplings $\{\zeta_D^{}, \rho_D^{}, 
\eta_{H_D}^{}, \eta_{\bar{H}_D}^{}\}$ replaced by $\{H_T^c, \bar{H}_T^c\}$ 
and $\{\zeta_T^{}, \rho_T^{}, \eta_{H_T}^{}, \eta_{\bar{H}_T}^{}\}$, 
respectively.  The terms in the first line of Eq.~(\ref{eq:HD-mass}) (and 
the corresponding ones for the Higgs triplets) give supersymmetric mass 
terms for $H_D$ and $\bar{H}_D$ (and for $H_T^c$ and $\bar{H}_T^c$). 
The masses for the fermionic components of the Higgs multiplets are 
then determined by the following equation:
\begin{equation}
  I^{(1)}_{H,R} I^{(1)}_{\bar{H},R} 
    - |\tilde{\zeta}_R^{}|^2 I^{(2)}_{H,R} I^{(2)}_{\bar{H},R} = 0,
\label{eq:KKmass-Higgsino}
\end{equation}
where $\tilde{\zeta}_R^{} \equiv -\zeta_R^{} \Lambda^{*2}/M_*^2$ and 
the index $R=D,T$ represents the doublet and triplet components.  The 
functions $I^{(1)}_{\Phi,R}$ and $I^{(2)}_{\Phi,R}$ are defined by
\begin{eqnarray}
  I^{(1)}_{\Phi,R} &=& 
    J_{c_\Phi\mp\frac{1}{2}}\Bigl(\frac{m_n}{k'}\Bigl) 
    - \frac{J_{c_\Phi\mp\frac{1}{2}}\bigl(\frac{m_n}{k}\bigl)}
    {Y_{c_\Phi\mp\frac{1}{2}}\bigl(\frac{m_n}{k}\bigl)}
    Y_{c_\Phi\mp\frac{1}{2}}\Bigl(\frac{m_n}{k'}\Bigl),
\label{eq:Idef-1} \\
  I^{(2)}_{\Phi,R} &=& 
    J_{c_\Phi\pm\frac{1}{2}}\Bigl(\frac{m_n}{k'}\Bigl) 
    - \frac{J_{c_\Phi\mp\frac{1}{2}}\bigl(\frac{m_n}{k}\bigl)}
    {Y_{c_\Phi\mp\frac{1}{2}}\bigl(\frac{m_n}{k}\bigl)}
    Y_{c_\Phi\pm\frac{1}{2}}\Bigl(\frac{m_n}{k'}\Bigl),
\label{eq:Idef-2}
\end{eqnarray}
where $\Phi = H, \bar{H}$, and $\mp$ ($\pm$) takes $-$ and $+$ ($+$ 
and $-$) for the doublets, $R=D$, and triplets, $R=T$, respectively. 
The scalar components receive contributions also from the terms in the 
second line in Eq.~(\ref{eq:HD-mass}) (and the corresponding ones for 
the triplets). Their masses are given by
\begin{eqnarray}
  && \Bigl( I^{(1)}_{H,R} I^{(1)}_{\bar{H},R} 
    - |\tilde{\zeta}_R^{}|^2 I^{(2)}_{H,R} I^{(2)}_{\bar{H},R} 
    + \tilde{\eta}_{H_{\!R}}^{} \frac{M'_*}{m_n} 
      I^{(1)}_{\bar{H},R} I^{(2)}_{H,R} \Bigr)
  \Bigl( I^{(1)}_{H,R} I^{(1)}_{\bar{H},R} 
    - |\tilde{\zeta}_R^{}|^2 I^{(2)}_{H,R} I^{(2)}_{\bar{H},R} 
    + \tilde{\eta}_{\bar{H}_{\!R}}^{} \frac{M'_*}{m_n} 
      I^{(1)}_{H,R} I^{(2)}_{\bar{H},R} \Bigr)
\nonumber\\
  && \qquad\qquad
    - |\tilde{\rho}_R^{}|^2 \frac{M'^2_*}{m_n^2} 
      I^{(1)}_{H,R} I^{(1)}_{\bar{H},R} 
      I^{(2)}_{H,R} I^{(2)}_{\bar{H},R} = 0,
\label{eq:KKmass-Higgsboson}
\end{eqnarray}
where $\tilde{\rho}_R^{} \equiv \rho_R^{} |\Lambda|^4/M_*^4$ and 
$\tilde{\eta}_{\Phi_{\!R}}^{} \equiv \eta_{\Phi_{\!R}}^{} |\Lambda|^4/M_*^4$. 
This equation reduces to Eq.~(\ref{eq:KKmass-Higgsino}) for 
$\tilde{\rho}_R^{} = \tilde{\eta}_{\Phi_{\!R}}^{} = 0$. In 
Eqs.~(\ref{eq:KKmass-Higgsino},~\ref{eq:KKmass-Higgsboson}) we 
have not included possible Planck-brane localized kinetic terms for $H_D$ 
and $\bar{H}_D$, $S = \int\!d^4x \int\!dy\, 2\delta(y) \int\!d^4\theta 
[(z_H/M_*) H_D^\dagger H_D + (z_{\bar{H}}/M_*) \bar{H}_D^\dagger \bar{H}_D]$.
Their effects are included by replacing $J_{c_\Phi-1/2}(m_n/k)$ and 
$Y_{c_\Phi-1/2}(m_n/k)$ in Eqs.~(\ref{eq:Idef-1},~\ref{eq:Idef-2}) 
with $J_{c_\Phi-1/2}(m_n/k)+(z_\Phi m_n/M_*)J_{c_\Phi+1/2}(m_n/k)$ 
and $Y_{c_\Phi-1/2}(m_n/k)+(z_\Phi m_n/M_*)Y_{c_\Phi+1/2}(m_n/k)$, 
respectively.  The corresponding terms for $H_T^c$ and $\bar{H}_T^c$ 
do not affect the spectrum because the triplet fields are localized 
to the TeV brane. 

It is useful to study the relative sizes of the masses obtained above. 
For small supersymmetry breaking $\Lambda \ll M_*$, we can expand 
Eqs.~(\ref{eq:KKmass-321gauginos},~\ref{eq:KKmass-XYfermion},~%
\ref{eq:KKmass-XYscalar},~\ref{eq:KKmass-Higgsino},~%
\ref{eq:KKmass-Higgsboson}) in powers of $\Lambda/M_*$. The masses 
of the lowest-lying modes in the gauge sector, $\lambda^{321}$, 
$\lambda'^{\rm XY}$ and $\sigma^{\rm XY}$, are then given by 
\begin{equation}
  m_{\lambda^{321}_a} = g_a^2 M_{\lambda,a}',
\end{equation}
\begin{equation}
  m_{\lambda'^{\rm XY}} = 2 g_B^2 k M_{\lambda,X}',
\end{equation}
\begin{equation}
  m_{\sigma^{\rm XY}}^2 = 2 g_B^2 k M_{\sigma,X}'^2,
\end{equation}
respectively, where $a=1,2,3$ represents the 321 gauge group. Here, 
$M_{\lambda,a}' \equiv M_{\lambda,a} e^{-\pi kR}$, $M_{\lambda,X}' 
\equiv M_{\lambda,X} e^{-\pi kR}$ and $M_{\sigma,X}' \equiv M_{\sigma,X} 
e^{-\pi kR}$ are parameters of order TeV, and $g_a \equiv (\pi R/g_B^2 
+ 1/\tilde{g}_{0,a}^2)^{-1/2}$ are the 4D gauge couplings.  Since $g_a = O(1)$ 
and $g_B^2 k = O(\pi kR)$ in the present theory, we find that the XY 
states, $\lambda'^{\rm XY}$ and $\sigma^{\rm XY}$, are expected to be 
heavier than the 321 gauginos, $\lambda^{321}_a$.  For instance, in the 
case of $M_{\lambda,a} \simeq M_{\lambda,X} \simeq M_{\sigma,X}/4\pi$, 
as suggested by naive dimensional analysis, the ratios of the masses are 
given by $m_{\lambda^{321}_a} : m_{\lambda'^{\rm XY}} : m_{\sigma^{\rm XY}} 
\simeq 1 : \pi kR : \pi kR$ (here we have regarded $\sqrt{\pi kR}$ to be 
$O(4\pi)$ in this estimate).  This can be understood in the 5D picture 
by the fact that the wavefunctions of the XY states are strongly localized 
to the TeV brane, where supersymmetry breaking occurs, while those of the 
321 gauginos are nearly conformally flat.  In the normalization where 
conformally flat modes have flat wavefunctions, the wavefunctions for 
$\lambda^{321}$, $\lambda'^{\rm XY}$ and $\sigma^{\rm XY}$ are roughly 
proportional to $1$, $e^{k|y|}$ and $e^{k|y|}$, respectively.  In the 4D 
dual picture the heaviness of the XY states arises because these states 
are composite states of the $G$ sector so that they interact more strongly 
with supersymmetry breaking caused by the dynamics of $G$. 

Similarly in the Higgs sector we can obtain the expressions for 
the masses in the small supersymmetry breaking limit.  For the 
doublet Higgs fields, the masses are expressed by the conventional 
set of parameters: the supersymmetric mass $\mu$, the holomorphic 
supersymmetry-breaking scalar mass $\mu B$, and non-holomorphic 
supersymmetry-breaking squared masses $m_\Phi^2$ ($\Phi = H, \bar{H}$). 
Defining 
\begin{equation}
  \delta_\Phi = \frac{1-e^{-(2c_\Phi-1)\pi kR}}{2c_\Phi-1}
    + \frac{z_\Phi k}{M_*},
\end{equation}
these parameters are given by
\begin{eqnarray}
  && \mu = \delta_H^{-1/2} \delta_{\bar{H}}^{-1/2} 
     e^{-(c_H-\frac{1}{2})\pi kR} e^{-(c_{\bar{H}}-\frac{1}{2})\pi kR}
     \tilde{\zeta}_D^{}\, k', 
\label{eq:HD-1} \\
  && \mu B = \delta_H^{-1/2} \delta_{\bar{H}}^{-1/2} 
     e^{-(c_H-\frac{1}{2})\pi kR} e^{-(c_{\bar{H}}-\frac{1}{2})\pi kR}
     \tilde{\rho}_D^{}\, k' M_*', 
\label{eq:HD-2} \\
  && m_\Phi^2 = \delta_\Phi^{-1} e^{-2(c_\Phi-\frac{1}{2})\pi kR} 
     \tilde{\eta}_{\Phi_D}^{}\, k' M_*'.
\label{eq:HD-3}
\end{eqnarray}
The masses for the triplet Higgs states are also specified by their 
supersymmetric mass, holomorphic supersymmetry-breaking scalar mass 
and non-holomorphic supersymmetry-breaking squared masses, which 
are denoted as $\mu_T^{}$, $\mu_T^{}\! B_T^{}$ and $m_{\Phi_T}^2$, 
respectively. Defining 
\begin{equation}
  \delta_{\Phi_T} = \frac{1-e^{-(2c_\Phi+1)\pi kR}}{2c_\Phi+1},
\end{equation}
they are given by
\begin{eqnarray}
  && \mu_T^{} = \delta_{H_T}^{-1/2} \delta_{\bar{H}_T}^{-1/2} 
     \tilde{\zeta}_T^{}\, k', 
\label{eq:HT-1} \\
  && \mu_T^{}\! B_T^{} = \delta_{H_T}^{-1/2} \delta_{\bar{H}_T}^{-1/2} 
     \tilde{\rho}_T^{}\, k' M_*', 
\label{eq:HT-2} \\
  && m_{\Phi_T}^2 = \delta_{\Phi_T}^{-1} 
     \tilde{\eta}_{\Phi_T}^{}\, k' M_*'.
\label{eq:HT-3}
\end{eqnarray}
Comparing Eqs.~(\ref{eq:HD-1}~--~\ref{eq:HD-3}) and 
Eqs.~(\ref{eq:HT-1}~--~\ref{eq:HT-3}), we find that the masses 
of the triplet states tend to be larger than those of the doublet 
states.  For instance, if all the dimensionless parameters are of the 
same order as expected in naive dimensional analysis, the ratios of 
the masses are given by $\mu/\mu_T^{} \simeq \mu B/\mu_T^{}\! B_T^{} 
\simeq m_\Phi^2/m_{\Phi_T}^2 = O(1/\pi kR)$ for $c_H \simeq c_{\bar{H}} 
\simeq 1/2$, and $\mu/\mu_T^{} \simeq \mu B/\mu_T^{}\! B_T^{} 
= O(e^{-(c_H+c_{\bar{H}}-1)\pi kR})$ and $m_\Phi^2/m_{\Phi_T}^2 
= O(e^{-(2c_\Phi-1)\pi kR})$ for $c_H, c_{\bar{H}} > 1/2$. 
This is again because the wavefunctions for the triplet 
states are localized to the TeV brane while those of the doublet 
states are not.  In the normalization where conformally flat 
modes have flat wavefunctions, the wavefunctions for the $H_D$, 
$\bar{H}_D$, $H_T^c$ and $\bar{H}_T^c$ states are roughly 
proportional to $e^{-(c_H-1/2)k|y|}$, $e^{-(c_{\bar{H}}-1/2)k|y|}$, 
$e^{(c_H+1/2)k|y|}$ and $e^{(c_{\bar{H}}+1/2)k|y|}$, respectively. 
In general, the GUT states such as the XY and triplet-Higgs states 
are naturally heavier than the MSSM states, because in the 4D picture 
they are composite states of $G$ and thus couple more strongly to the 
supersymmetry breaking than the elementary states.  This is strongly 
related to the fact that we obtain the MSSM prediction for gauge 
coupling unification, which arises because the GUT states are composite 
particles so that their contribution to the gauge coupling evolution 
shuts off above the compositeness scale of order TeV. 

So far we have only considered the masses generated at tree level in 
5D.  The MSSM squarks and sleptons and $A_5^{\rm XY}$ are still massless 
at this level.  They, however, obtain masses at one loop through the 
standard model gauge interactions.  Because of the geometrical separation 
between supersymmetry breaking and the place where squarks and sleptons 
are located, the generated squark and slepton masses are finite and 
calculable in the effective field theory.  The calculation of these 
masses has been carried out in Ref.~\cite{Nomura:2003qb} and the result 
is given by
\begin{equation}
  m_{\tilde{f}}^2 = 
    \frac{1}{2\pi^2}\! \sum_{a=1,2,3}\! C_a^{\tilde{f}}\, {\cal I}_a,
\label{eq:MSSMscalar}
\end{equation}
where $\tilde{f} = \tilde{q}, \tilde{u}, \tilde{d}, \tilde{l}, \tilde{e}$ 
represents the MSSM squarks and sleptons, and the $C_a^{\tilde{f}}$ are 
the group theoretical factors given by $(C_1^{\tilde{f}}, C_2^{\tilde{f}}, 
C_3^{\tilde{f}}) = (1/60,3/4,4/3)$, $(4/15,0,4/3)$, $(1/15,0,4/3)$, 
$(3/20,3/4,0)$ and $(3/5,0,0)$ for $\tilde{f} = \tilde{q}, \tilde{u}, 
\tilde{d}, \tilde{l}$ and $\tilde{e}$, respectively.  The functions 
${\cal I}_a$ are defined in Eq.~(21) of~\cite{Nomura:2003qb}. Because 
these masses are generated through  gauge interactions, they are 
flavor universal and the supersymmetric flavor problem is absent. 
Another important point is that we have to use three different gaugino 
mass parameters $M_{\lambda,a}$ in the functions ${\cal I}_a$ because 
of the non-universal gaugino masses.  This has interesting consequences 
on the phenomenology of the model as discussed in the next section. 

The mass of $A_5^{\rm XY}$ is similarly generated at one loop through 
gauge interactions.  It receives contributions from loops of the 321 
gauginos as well as those of the 321 gauge bosons, with the former cutting 
off the quadratically divergent contribution from the latter.  The remaining 
logarithmic divergence is then made finite at the scale $\pi k'$ by the 
loops of the KK towers.  This is because in the 4D picture $A_5^{\rm XY}$ 
is the pseudo-Goldstone boson associated with the breaking of the global 
symmetry of $G$, $SU(5) \rightarrow {\rm 321}$, which is encoded as the 
boundary condition breaking {\it at the TeV brane}.  This implies that 
$A_5^{\rm XY}$ can receive a mass only through the explicit breaking of 
the global $SU(5)$ symmetry of $G$, which corresponds to the boundary 
condition breaking {\it at the Planck brane}.  The $A_5^{\rm XY}$ mass 
is then finite and calculable, since $A_5^{\rm XY}$ (and supersymmetry 
breaking) is localized to the TeV brane so that the loops generating 
the $A_5^{\rm XY}$ mass must probe both the Planck and the TeV branes. 
We thus find that the mass of $A_5^{\rm XY}$ is given by 
\begin{equation}
  m_{A_5^{\rm XY}}^2 \simeq \frac{g^2 C}{\pi^4}
    m_{\lambda'^{\rm XY}}^2 \ln\frac{\pi k'}{m_{\lambda'^{\rm XY}}},
\end{equation}
where $g$ represents a 4D gauge coupling and $C$ the group theoretical 
factor.  This expression is valid as long as $m_{\lambda'^{\rm XY}} 
\simgt m_{\lambda^{321}_a}$, which is expected to be the case.  Since 
$A_5^{\rm XY}$ is charged under $SU(3)_C$, we should take $g = g_3$ 
and $C = 4/3$ in the above estimate. 

To summarize, we have seen that supersymmetry breaking at the TeV brane 
naturally gives all the masses necessary to make the model viable, 
with the sizes of the masses depending on the states.  It is useful 
to classify the fields into several categories to figure out the 
relative sizes of their masses.  We can first divide the fields into 
two categories, the fields regarded respectively as  (A) elementary 
and (B) composite fields in the 4D picture.  The former consists of the 
MSSM states, $\lambda^{321}, H_D, \bar{H}_D$ and $\tilde{f}$, while the 
latter consists of the GUT states, $\lambda'^{\rm XY}, \sigma^{\rm XY}, 
A_5^{\rm XY}, H_T^c$ and $\bar{H}_T^c$.  In the 5D picture class (B) 
corresponds to the states which are localized to the TeV brane, while (A) 
corresponds to the states which are not.  Since supersymmetry breaking 
is caused by the dynamics of $G$ (localized to the TeV brane in the 5D 
picture) the states in (B) generically receive larger masses than those 
in (A).  We could also divide the fields into the classes which receive 
masses at (I) tree and (II) loop levels in 5D.  The class (I) contains 
$\lambda^{321}, \lambda'^{\rm XY}$, $\sigma^{\rm XY}$, $H_T^c$ and 
$\bar{H}_T^c$ (and $H_D, \bar{H}_D$ for $c \simeq 1/2$), while the class 
(II) contains $\tilde{f}$ and $A_5^{\rm XY}$ (and $H_D, \bar{H}_D$ for 
$c \gg 1/2$).  The masses of the fields in (II) are naturally suppressed 
compared with those in (I) by a loop factor (but the logarithm between 
the masses and the KK scale could make this hierarchy small, especially 
when supersymmetry breaking is weak). The fields belonging to the four 
classes (A-I), (A-II), (B-I) and (B-II) are depicted explicitly in 
Table~\ref{table:class}, and the general trend for the spectrum in the 
gauge sector is depicted in Fig.~\ref{fig:spectrum}b.
\begin{table}
\begin{center}
\begin{tabular}{|c|c|c|}
\hline
 & (A) ``elementary'' & (B) ``composite'' \\ \hline
 (I) tree  & $\lambda^{321}$, \{$H_D, \bar{H}_D$ for $c \simeq 1/2$\} & 
    $\lambda'^{\rm XY}, \sigma^{\rm XY}, H_T^c, \bar{H}_T^c$ \\ 
 (II) loop & $\tilde{f}$, \{$H_D, \bar{H}_D$ for $c \gg 1/2$\} & 
    $A_5^{\rm XY}$ \\
\hline
\end{tabular}
\end{center}
\caption{The classification of the fields obtaining masses from 
 supersymmetry breaking; see the text.}
\label{table:class}
\end{table}
Assuming that coefficients of all the operators scale according to 
naive dimensional analysis, the ratios of the masses for fields in 
each category are estimated roughly as $m_{\rm A-I}^{} : m_{\rm A-II}^{} 
: m_{\rm B-I}^{} : m_{\rm B-II}^{} \approx 1 : g_B^2 k/16\pi^2 : \pi kR 
: (g^2 C/\pi^4)^{1/2} \pi kR$. We should, however, emphasize that the 
operators giving the masses for these states have coefficients which are free 
parameters (see e.g. Eqs.~(\ref{eq:gaugino-mass},~\ref{eq:XYfermion-mass},~%
\ref{eq:XYscalar-mass},~\ref{eq:HD-mass})).  In particular, the operators 
giving masses for $\lambda'^{\rm XY}$ and $\sigma^{\rm XY}$ involve the 
$y$ derivative so that they may have somewhat suppressed coefficients. 
Therefore, it could well be possible that some of these states, for example 
$\lambda'^{\rm XY}$, $\sigma^{\rm XY}$ and $A_5^{\rm XY}$, are lighter than 
the naive estimate given above, improving the prospects for their discovery 
at future colliders.

\section{Phenomenology}
\label{sec:pheno}

For a theory that incorporates something that closely resembles MSSM 
gauge coupling unification, the spectrum of MSSM particles in the present 
model can be quite unusual.  Because the gaugino masses originate from 
supersymmetry breaking terms on the $SU(5)$-violating TeV brane, there 
is no guarantee that the gluino will be much heavier than the wino, which 
will in turn be heavier than the bino.  Moreover, the squark and slepton 
masses are generated radiatively through gaugino loops, as determined 
by Eq.~(\ref{eq:MSSMscalar}), so the scalar spectrum inherits whatever 
unusual features distinguish the gaugino spectrum.

To obtain the physical masses of the gauginos and scalars, the values 
given by Eqs.~(\ref{eq:KKmass-321gauginos},~\ref{eq:MSSMscalar}), which 
are the running masses at the KK scale $m_{\rm KK}^{} \simeq \pi k'$, 
must be run down in energy. For the scalar masses-squared, the $D$-term 
contributions $\Delta=(T_3-Q \sin^2\!\theta_W) \cos 2\beta \; m_Z^2$ must 
also be included.  The smaller the size of supersymmetry breaking on the 
TeV brane, the larger the energy interval between $m_{\rm KK}^{}$ and 
the superpartner masses, and the larger the running effect will be. 
For very strong supersymmetry breaking, the squark and slepton masses 
will be considerably smaller than the gaugino masses, while for weaker 
supersymmetry breaking masses, the scalar and gaugino masses can be 
comparable.  This is evident in the plots of~\cite{Nomura:2003qb}, where 
with $x \equiv M_\lambda/k$, the ratio of the TeV-brane gaugino mass to 
the curvature scale, ranging from 10 to 0.01, the ratio of the bino mass 
to the right-handed selectron mass falls from roughly a factor of eight 
to less than a factor of two. 

One interesting consequence of the non-universality of the TeV-brane 
gaugino masses is that there are many possibilities for which particle 
is the next-to-lightest supersymmetric particle (NLSP) (the NLSP decays 
promptly into the LSP gravitino, whose mass is $\sim k'^2/M_{\rm Pl} 
\sim 0.01-0.1~{\rm eV}$).   If the TeV-brane gaugino masses are universal 
as in~\cite{Nomura:2003qb}, the NLSP will be the right-handed stau. 
More generally, this is the result when the dominant contributions 
to the scalar masses are from gluino and wino loops -- in the notation 
of Eq.~(\ref{eq:MSSMscalar}), ${\cal I}_3,\; {\cal I}_2>{\cal I}_1$. 
If instead ${\cal I}_3,\; {\cal I}_1>{\cal I}_2$, it is possible for 
the bino-loop contribution to $m^2_{\tilde{e}}$ to be larger than the 
wino-loop contribution to $m^2_{\tilde{l}}$, giving a sneutrino NLSP. 
A right-handed bottom squark NLSP is possible if ${\cal I}_2,\; 
{\cal I}_1>{\cal I}_3$, and even the left-handed stop could technically be 
the NLSP, although it requires ${\cal I}_1 \gg\; {\cal I}_2,\; {\cal I}_3$. 
In fact, since all of the scalars get contributions from bino loops, one 
can even have scenarios with either wino or gluino NLSP's.  Some of these 
cases are likely to be more natural than others from the perspective of 
electroweak symmetry breaking; here we have simply enumerated some of 
the various possibilities.

In the limit that the squarks and sleptons are heavy enough that $D$-term 
splittings are not very important, the three quantities ${\cal I}_1$, 
${\cal I}_2$, and ${\cal I}_3$ determine the five masses for $\tilde{q}$, 
$\tilde{u}$, $\tilde{d}$, $\tilde{l}$, and $\tilde{e}$.  If the gluino 
contribution dominates the masses of $\tilde{q}$, $\tilde{u}$, and 
$\tilde{d}$, they will be quite degenerate, and only by resolving this 
degeneracy can the sum rules 
\begin{eqnarray}
  m^2_{\tilde{q}} &=& 
    m^2_{\tilde{d}} + m^2_{\tilde{l}} - \frac{1}{3} m^2_{\tilde{e}}, \\
  m^2_{\tilde{u}} &=& 
    m^2_{\tilde{d}} + \frac{1}{3} m^2_{\tilde{e}},
\end{eqnarray}
be tested. On the other hand, if ${\cal I}_1$, ${\cal I}_2$, and ${\cal I}_3$ 
are all comparable, there will not be a strong hierarchy among the scalar 
masses, and the above relations will be more easily tested.  Such a scenario 
requires the brane mass for the gluino to be somewhat smaller than that for 
the wino, for instance.  In reality, it is quite possible that $D$-term 
contributions will be important, in which case the above sum rules will be 
modified.  In that case, however, there are four parameters (now including 
$\tan\beta$) that fix seven scalar masses, so the more general point still 
remains --- the model predicts relations among the scalar masses, and these 
will be most easily tested if the scalar masses are all comparable in size.%
\footnote{For the third generation, the relations described here will be 
modified by Yukawa effects.}

A very predictive spectrum arises in the limit of very strong supersymmetry 
breaking on the TeV brane, in which case the entire mass spectrum of 
the theory is essentially determined by the single free parameter $k'$. 
This spectrum for the MSSM gauginos and scalars is identical to the one 
described in~\cite{Nomura:2003qb} for $x \gg 1$.  A distinctive feature 
is that the MSSM gauginos combine with the conjugate gauginos from 
$\Sigma^{321}$ to form pseudo-Dirac states in this limit.  An important 
difference compared to~\cite{Nomura:2003qb} is that here the XY gauginos 
do not become massless in the limit of infinitely strong supersymmetry 
breaking.  Instead, $\lambda^{\rm XY}$ and $\lambda'^{\rm XY}$ appear 
in Dirac states, the lightest of which are nearly degenerate with the 
lightest KK excitations of the standard-model gauge bosons, $m \simeq 
(3\pi/4) k'$.  The lightest $\sigma^{\rm XY}$ mode also has this mass, 
as can be verified using Eq.~(\ref{eq:KKmass-XYscalar}).  The mass of 
the lightest $A_5^{\rm XY}$ mode is of order $k'/\pi$.

As discussed in section~\ref{sec:susy-breaking}, the lightest of the 
GUT particles (LGP) is likely to be $A_5^{\rm XY}$ because its mass is 
loop suppressed.  The LGP, which is generally colored, will be stable 
at least for collider purposes.  This is because the LGP is localized 
to the TeV brane, so that its decay to quarks is highly suppressed 
(there is no bulk mode that can carry a color charge from the TeV 
brane to the Planck brane at a sufficiently large rate).%
\footnote{The GUT parity of~\cite{Goldberger:2002pc} 
can be badly broken on the TeV brane in the present model.}
Since the LGP is colored, it hadronizes after production.  These exotic 
hadrons are charged or neutral, depending on whether the LGP picks 
up an up or down quark or antiquark, but the mass differences among 
them are small enough that they are both stable for collider purposes. 
The charged ones would thus be detectable at collider experiments 
through highly ionizing tracks~\cite{Goldberger:2002pc}.  In the case 
that $A_5^{\rm XY}$ is the LGP, the exotic hadrons are four fermionic 
mesons: $\tilde{T}^0$, $\tilde{T}^-$, $\tilde{T}'^-$ and $\tilde{T}^{--}$ 
(and their anti-particles).  We can estimate the reach of the LHC to be 
roughly $2~{\rm TeV}$ in the masses of these states~\cite{Rizzo:1996ry}.

Finally, we make a few remarks about the Higgs sector in the model.  If 
the Higgs multiplets are strongly localized to the Planck brane, their 
soft masses-squared are generated radiatively just as for the other MSSM 
scalars.  At one loop, wino and bino loops give a positive contribution 
equal to the one-loop contribution to $m^2_{\tilde{l}}$.  A two-loop 
negative contribution to $m_{H}^2$ coming from the top Yukawa interaction 
will overcome this positive contribution provided the gluino mass is 
not too small.  In this case of brane-localized Higgs fields, some 
mechanism for generating $\mu$ must be introduced.  One possibility 
is to generate $\mu$ by ``shining'' it from the TeV brane, as described 
in~\cite{Goldberger:2002pc}; $\mu B$ can then be generated radiatively 
through gaugino loops unless the supersymmetry breaking on the TeV brane 
is so strong that the gauginos are essentially Dirac in nature.  If 
the Higgs fields are delocalized, then $\mu$, $\mu B$, $m_{H}^2$, and 
$m_{\bar{H}}^2$ all arise at tree level.  For these parameters to be 
generated with the appropriate size for natural electroweak symmetry 
breaking, the Higgs hypermultiplet mass parameters should be close to 
their sizes in the conformal limit: $c_H \sim c_{\bar{H}} \simeq 0.5-0.6$ 
(see Eqs.~(\ref{eq:HD-1}~--~\ref{eq:HD-3})).

\section{Conclusions}
\label{sec:concl}

We have studied a model of warped supersymmetric unification in which 
neither the Planck brane nor the TeV brane respects the $SU(5)$ unified 
symmetry.  Before supersymmetry breaking, the massless spectrum contains 
fermions and scalars with the quantum numbers of XY gauge fields, in 
addition to the particle content of the MSSM.  In the 4D description, 
these states make up the pseudo-Goldstone supermultiplets associated 
with the spontaneous breaking of the approximate $SU(5)$ global symmetry. 
We have shown that these states can acquire masses from supersymmetry 
breaking terms on the TeV brane, and that they leave intact the unification 
prediction for the low-energy gauge couplings.  Whether they will be 
accessible to future colliders such as the LHC depends on the sizes of 
coefficients of operators that produce their masses (e.g. $\eta$ and 
$\rho$ in Eqs.~(\ref{eq:XYfermion-mass},~\ref{eq:XYscalar-mass})). 
The lightest of these states are expected to be the scalars 
$A_5^{\rm XY}$, which have loop-suppressed masses. 

The model we have presented leads to a spectrum of MSSM gauginos and 
scalars that can be quite unusual.  This stems from the fact that 
these particles acquire mass (either at tree level or radiatively) from 
supersymmetry breaking terms on the TeV brane, where the unified symmetry 
is not realized.  Consequently, while the model is valid as an effective 
theory up to very high scales and incorporates gauge coupling unification, 
aspects of the spectrum that are essentially fixed in more conventional 
supersymmetric unification --- for example, the ordering of the gaugino 
masses --- are here allowed a range of possibilities.

\section*{Acknowledgment}

The work of Y.N. and B.T. was supported in part by the Director, Office 
of Science, Office of High Energy and Nuclear Physics, of the U.S. 
Department of Energy under Contract DE-AC03-76SF00098.

\end{document}